%% file: main.tex
\RequirePackage{lineno}
\documentclass[aps,twocolumn,showpacs,byrevtex,prl,reprint]{revtex4-1}

\usepackage{graphicx}
\usepackage{dcolumn}
\usepackage{bm}
\usepackage{rotating}
\usepackage{epstopdf}
\usepackage{color}
\usepackage{verbatim} 
\usepackage{multirow}
\usepackage[abs]{overpic}
\usepackage{subfigure}
\usepackage{xspace}
\usepackage{tikz}
\usepackage{amsthm,amsmath,amssymb}
\usepackage{mathrsfs}
\usepackage[normalem]{ulem}
\usepackage{lineno}
\usepackage{hyperref}
\usepackage[hyphenbreaks]{breakurl}
\usepackage{xurl}

\newcommand{\PreserveBackslash}[1]{\let\temp=\\#1\let\\=\temp}
\newcolumntype{C}[1]{>{\PreserveBackslash\centering}p{#1}}
\newcolumntype{R}[1]{>{\PreserveBackslash\raggedleft}p{#1}}
\newcolumntype{L}[1]{>{\PreserveBackslash\raggedright}p{#1}}
\newcommand{\pp}{\pi^+\pi^-}

\newcommand{\EE}{e^+e^-}

\newcommand{\GG}{\gamma\gamma}
\newcommand{\psip}{\psi(3686)}

\newcommand{\jpsi}{J/\psi}
\newcommand{\chicJ}{\chi_{cJ}}

\let\oldequation\equation
\let\oldendequation\endequation
\renewenvironment{equation}{\linenomathNonumbers\oldequation}{\oldendequation\endlinenomath}
\begin{document}
\graphicspath{{figure/}}
\DeclareGraphicsExtensions{.eps,.png,.ps}
\title{\boldmath Observation of \texorpdfstring{$\eta_{c}$}{eta_{c}}(1S, 2S) and \texorpdfstring{$\chi_{cJ}$}{chi_{cJ}} decays to 2(\texorpdfstring{$\pi^{+}$}{pi^{+}}\texorpdfstring{$\pi^{-}$}{pi^{-}})\texorpdfstring{$\eta$}{eta} via \texorpdfstring{$\psi$}{psi}(3686) radiative transitions}
\input{authors}



\begin{abstract}
Based on $2.7 \times 10^9~\psi(3686)$ decays collected with the BESIII detector,
the radiative decay $\psi(3686)\to\gamma2(\pi^{+}\pi^{-})\eta$ is investigated to measure properties of S- and P-wave charmonium states.
The branching fraction of the decay $\eta_{c}(1S) \to 2(\pi^{+}\pi^{-})\eta$, which is found to have a strong dependence on the interference pattern between $\eta_c(1S)$ and non-$\eta_c(1S)$ processes, is measured in both destructive and constructive interference scenarios for the first time. The mass and width of the $\eta_{c}(1S)$ are measured to be $M=(2984.14 \pm 0.13 \pm 0.38)$ MeV/$c^{2}$ and $\Gamma=(28.82 \pm 0.11 \pm 0.82)$ MeV, respectively. Clear signals for the decays of the $\chi_{cJ}(J=0,1,2)$ and the $\eta_{c}(2S)$ to $2(\pi^{+}\pi^{-})\eta$ are also observed for the first time, and the corresponding branching fractions are measured. The ratio of the branching fractions between the $\eta_{c}(2S)$ and $\eta_{c}(1S)$ decays is significantly lower than the theoretical prediction, which might suggest different dynamics in their decays.
\end{abstract}

\maketitle

Our understanding of charmonium states with masses below the open-charm production threshold is far from satisfactory.  This is especially true for the spin singlet states~\cite{Ref_a}, which include the S-wave ground state $\eta_{c}$ ($\eta_{c} \equiv \eta_{c}(1S)$) and its first radial excitation $\eta_{c}(2S)$. The $\eta_{c}(2S)$ state was first observed by the Belle experiment~\cite{Ref_b} in $B$ meson decay and later confirmed by the BaBar~\cite{Ref_c} and CLEO \cite{Ref_d} experiments. BESIII was the first to observe the radiative transition $\psip \to \gamma\eta_{c}(2S)$~\cite{Ref_g}. The sum of the measured branching fractions (BFs) of the $\eta_{c}(2S)$ is currently only about 6$\%$. Many decay modes of the $\eta_c$ also remain unknown and some measured BFs have large uncertainties. 

The expected ratio of BFs for decays of the $\eta_{c}$ and $\eta_{c}(2S)$ into the same hadronic final state was initially anticipated to be similar to the ratio (12$\%$) observed for their spin-triplet partners $\jpsi$ and $\psi(3686)$~\cite{Ref_09,Ref_06}. However, a more in-depth analysis suggested that this ratio should actually be close to unity, as the dominant decay dynamics of the $\eta_{c}$ and $\eta_{c}(2S)$ involve the annihilation of charmed quark pairs into two gluons~\cite{Ref_10}. Experimental measurements have revealed inconsistencies with both of these theoretical predictions~\cite{Ref_11}. Further measurements of additional decay modes of the $\eta_{c}$ and $\eta_{c}(2S)$, in addition to  testing these predictions, may provide insight into the internal structure of these states, for example by testing for the presence of non-charmonium components in their wave functions.

Previous measurements of the $\eta_c$ mass and width, obtained through the magnetic dipole~(M1) transition $\psi(3686)\to\gamma\eta_c$, $p\bar{p}$ collisions, or $\gamma\gamma$ collisions,  have shown significant discrepancies between different experiments~\cite{Ref_02}. However, interference between $\eta_c$ decays and non-resonant processes could have a large impact on these measurements as well as on $\eta_c$ BFs. The CLEO~\cite{Ref_04}, BESIII~\cite{Ref_05}, and KEDR~\cite{Ref_03} collaborations have studied $\eta_c$ decays using the M1 transition $\psi(3686)\to\gamma\eta_c$, which is considered to be a golden channel for $\eta_c$ production due to its low background level and the few unknown background channels. BESIII successfully described the $\eta_c$ line shape by using the energy dependence of the hindered-M1 transition matrix element and by fully considering interference with non-resonant $\psi(3686)$ radiative decays, leading to a significant improvement in the precision of the $\eta_c$ mass and width measurements~\cite{Ref_05}.  However, further improvements are still beneficial since, for example, the determination of the hyperfine mass splitting of S-wave charmonium can test theoretical calculations~\cite{Ref_hf1s1,Ref_hf1s2}, and the width directly relates to the theory of $\eta_c$-glueball mixing~\cite{cheny}.

As for the $P$-wave spin-triplet states $\chi_{c1,2}$, the dominant decay modes are the transitions $\chi_{c1,2}\to\gamma\jpsi$. However, the sum of all measured $\chi_{c0,1,2}$ BFs are each still far less than 1. Intensive studies of their multi-body decays are lacking relative to their few-body decays. The search for more new decay modes of the $\chi_{c0,1,2}$ is useful in understanding their properties.

In this Letter, we analyze $2.7 \times 10^9~\psi(3686)$ decays to study the radiative decay $\psi(3686)\to \gamma 2(\pp)\eta$.  
We report the first observations of $\eta_{c}(2S)$ and $\chi_{cJ}$ decays to $2(\pp)\eta$. We also present improved measurements of the $\eta_{c}$ mass and width and the BF of $\eta_c \to 2(\pp)\eta$. We find that the BF strongly depends on the interference between the $\eta_{c}$ and non-$\eta_{c}$ components.

The design and performance of the BESIII detector are described in Refs~\cite{besiii,besiii2}. The corresponding simulation, analysis framework, and software are presented in Refs~\cite{evtgen1,evtgen2}. Simulated Monte Carlo (MC) samples are produced with {\sc geant4}-based~\cite{G4} software, which models the experimental conditions, including the electron-positron collision, the decays of the particles, and the response of the detector. Final-state radiation (FSR) from charged final-state particles is incorporated using the {\sc photos} package~\cite{fsr}. The exclusive decays of $\psi(3686) \to \gamma X$ are generated following the angular distribution of (1 + $\lambda \cos^{2}\theta_{\gamma}$), where $X$ refers to $\eta_{c}(1S, 2S)$ or $\chi_{cJ}$, $\theta_{\gamma}$ is the polar angle of the radiative photon in the rest frame of the $\psi(3686)$, and the value of $\lambda$ is set to 1 for $\eta_{c}(1S, 2S)$ and to 1, -1/3, 1/13 for $\chi_{c0}$, $\chi_{c1}$, and $\chi_{c2}$ respectively~\cite{Ref_12, Ref_05}. The $X \to 2(\pp)\eta$ decays are generated uniformly in phase space (PHSP).

Charged tracks detected in the multilayer drift chamber (MDC) are required to be within the polar angle ($\theta$) range of $|\cos\theta|<0.93$. Here $\theta$ is defined with respect to the $z$-axis, which is the symmetry axis of the MDC. For each charged track, the distance of closest approach to the interaction point must be less than 10 cm along the $z$-axis and less than 1 cm in the transverse plane. Photon candidates are reconstructed from isolated showers in the electromagnetic calorimeter (EMC). The deposited energy of each shower is required to be at least 25 MeV in the barrel region $(|\cos\theta|<0.80)$ and 50 MeV in the end cap region $(0.86<|\cos\theta|<0.92)$. For the $\eta_{c}(2S)/\chi_{cJ}$ modes, the deposited energy is required to be larger than 40 MeV in the end cap region due to the low energy of the M1 transition photon. To exclude showers originating from charged tracks, the angle between the shower direction and the charged tracks extrapolated to the EMC must be greater than $10^\circ$. To suppress electronic noise and energy depositions unrelated to the event, the EMC cluster timing from the reconstructed event start time is required to be within [0, 700] ns.

Candidate events are required to have four charged tracks with zero net charge and at least three photons. All the charged tracks are taken as pions. To reconstruct the $\eta\to\gamma\gamma$ decay, the invariant mass of a pair of photons is required to satisfy $M_{\gamma\gamma}\in$[0.51, 0.57]~GeV/$c^2$. Then a one-constraint (1C) kinematic fit is performed for each combination with $\chi^{2}_{\rm{1C}}(\gamma\gamma)<20$, where the photon pair is constrained to the $\eta$ mass, and at least one $\eta$ candidate is required. To suppress background and improve the mass resolution, a five-constraint (5C) kinematic fit is performed to the total initial four momentum of the colliding beams. The extra 1C is used to constrain the $M_{\GG}$ to the $\eta$ nominal mass. If there is more than one combination in an event, the one with the minimum $\chi^{2}$ value of the 5C kinematic fit ($\chi^{2}_{\rm{5C}}$) is chosen. Furthermore, the criterion $\chi^{2}_{\rm{5C}}<15$ is required, which is determined by optimizing the figure of merit, defined as $S/\sqrt{S+B}$, where $S (B)$ is the number of signal (background) events in the fit region. In addition, to suppress backgrounds from processes with missing or additional photons, a four-constraint (4C) kinematic fit is performed for both signal and  background channels. For the $\gamma\eta_{c}$ mode, the $\chi^{2}_{\rm{4C}}(3\gamma2(\pp))$ is required to be less than $\chi^{2}_{\rm{4C}}(4\gamma2(\pi^{+}\pi^{-}))$ and $\chi^{2}_{\rm{4C}}(2\gamma2(\pi^{+}\pi^{-}))$. For the $\gamma\eta_{c}(2S)/\chi_{cJ}$ modes, only $\chi^{2}_{\rm{4C}}(3\gamma2(\pp))<\chi^{2}_{\rm{4C}}(4\gamma2(\pi^{+}\pi^{-}))$ is required since the events with one missing photon mainly come from FSR, which will be discussed later. Backgrounds from the continuum process $\EE \to q\bar{q}$ are investigated using the off-resonance data sample taken at the center-of-mass (CM) energy of 3.65 GeV. It is found that its shape is generally smooth in the $\eta_c(1S,2S)$ and $\chicJ$ mass regions. 

Background events from the $\psi(3686) \to P\jpsi (P=\pp,\eta)$ process are suppressed by requiring the recoil mass of $P$ to not fall in the $\jpsi$ signal region. In case of multiple $\pp$ combinations in an event, we select the one with $M^{\rm recoil}_{P}$ closest to the $\jpsi$ mass. Background events from the $\psi(3686) \to \pi^{0} H$ process ($H$ denotes hadronic states) are suppressed by requiring the invariant mass of all possible photon pair combinations to be outside the $\pi^0$ signal region. For $\gamma\eta_{c}(2S)/\chi_{cJ}$ decay modes, the number of background events from the $\psi(3686) \to \gamma\chi_{cJ}$; $\chi_{cJ} \to \gamma\jpsi$ process are suppressed by requiring the recoil invariant mass of all possible photon pair combinations to not fall in the $\jpsi$ signal region.

The process $\psi(3686) \to 2(\pi^{+}\pi^{-})\eta$ can contaminate our signal channel through FSR from one of the final-state pions.  This effect is particularly notable in the $\eta_{c}(2S)$ mass region, where the FSR photon has a similar energy to the radiative photon. To separate this background from the $\eta_{c}(2S)$ signal, we perform a modified kinematic fit where the energy of the radiative photon is floating in the 5C fit (m5C). Although the resolution of the $\eta_c(2S)$ signal itself becomes slightly worse, the FSR background shape becomes smoother. This results in a widened gap between the $\eta_c(2S)$ peak and FSR background~\cite{Ref_12}. Furthermore, the background shape is sensitive to the fraction of the number of background events between FSR and non-FSR modes. The discrepancy between data and MC simulation is corrected using a control sample $\psi(3686) \to \gamma\chi_{c0},\chi_{c0} \to 3(\pi^{+}\pi^{-})(\gamma_{\rm{FSR}})$ using the same method as in Ref.~\cite{Ref_12}. 

The process $\psi(3686) \to \pi^{0}2(\pi^{+}\pi^{-})\eta$, where a photon from the $\pi^0$ decay is lost, is the dominant remaining background. Particularly in the $\eta_c$ mass region, a data-driven method is employed to determine its contribution by comparing the detection efficiencies between the signal and background channels, and combining with the actual number of events in data~\cite{Ref_05}. 

After a topology analysis with the $\psi(3686)$ inclusive MC events~\cite{Ref_topo}, we find that there are several potential peaking backgrounds for the $\chicJ$ modes. The normalized contributions for these backgrounds using branching fractions from the PDG~\cite{Ref_02} are listed in Table~\ref{potential}, and we will subtract them from the signals directly when calculating the BFs.

\begin{table}[htp]
\begin{center}
\caption{Estimates of potential peaking backgrounds.}
\label{potential}
\begin{tabular}{c c c c}
\hline

        Backgrounds & $\chi_{c0}$  & $\chi_{c1}$  & $\chi_{c2}$ \\ \hline
        $\chi_{c0} \to \eta^{\prime}\eta^{\prime};\eta^{\prime} \to \gamma\pi^{+}\pi^{-}$  &  287$\pm$17  &  0  &  0    \\\hline
        $\chi_{cJ} \to \gamma\jpsi;\jpsi \to 2(\pi^{+}\pi^{-})\pi^{0}$  &  0  & 210$\pm$21  &  85$\pm$9   \\\hline
        $\chi_{cJ} \to \gamma\jpsi;\jpsi \to \gamma2(\pi^{+}\pi^{-})$  &  0  &  55$\pm$10  &  2.0$\pm$0.4   \\
         \hline

\end{tabular}
\end{center}
\end{table}

\begin{figure}[htb]
\begin{center}
\includegraphics[width=.44\textwidth]{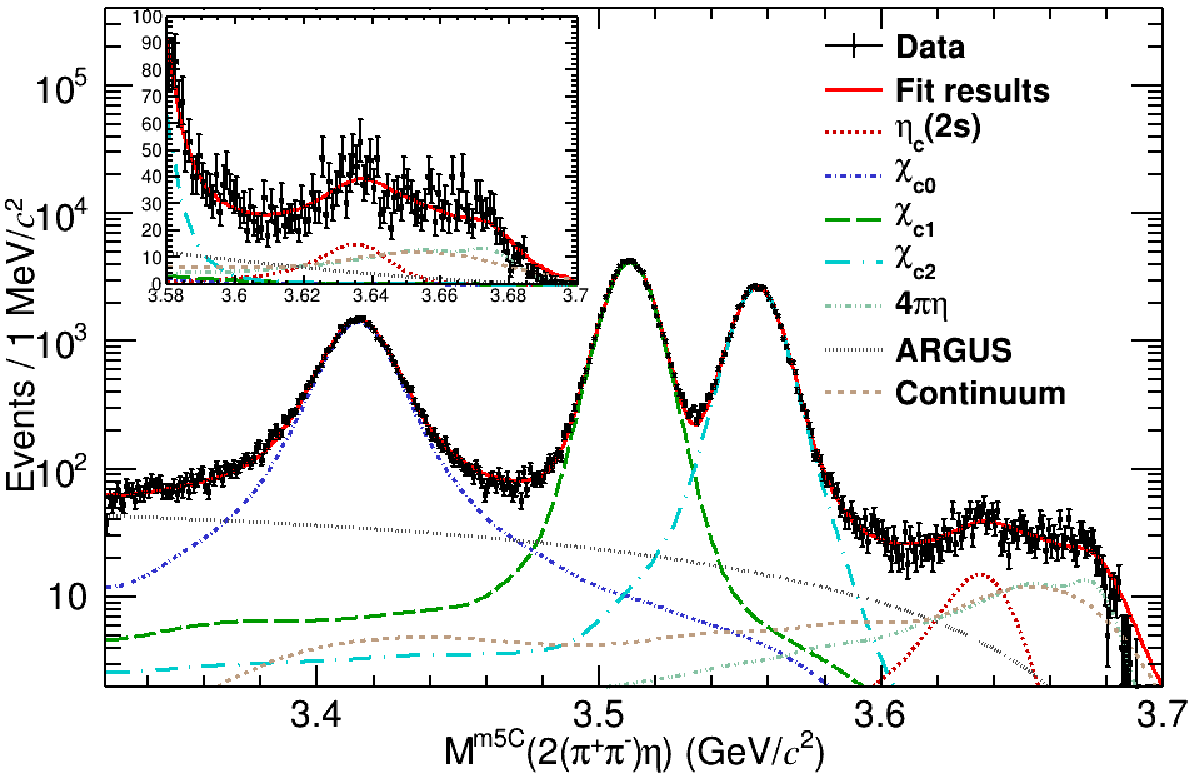}
\caption{Fit to the $M^{m5C}_{2(\pi^{+}\pi^{-})\eta}$ spectrum. The different components are expressed with different line styles or colors as given in the legend.}
\label{2sfit}
\end{center}
\end{figure}

The signal yields for $\chi_{cJ}/\eta_{c}(2S)$ are obtained by performing an unbinned maximum likelihood fit to the $M_{2(\pi^{+}\pi^{-})\eta}^{m5C}$ distribution as shown in Fig.~\ref{2sfit}. The fit function includes the $\chi_{cJ}/\eta_{c}(2S)$ signals, continuum background, FSR background from $\psi(3686) \to 2(\pi^{+}\pi^{-})\eta$, and other background contributions from $\psi(3686)$ decays. The $\chi_{cJ}$ signal functions are described by MC-simulated shapes convolved with Gaussian functions to account for the difference in detector resolution between data and MC simulation, and the parameters of the smeared Gaussian functions are free. The $\eta_{c}(2S)$ signal function is described by
\begin{equation}
\label{lineshape}
\textstyle\begin{footnotesize}{\epsilon(m)(E_{\gamma}^{3} \times {BW(m)} \times f_{d}(E_{\gamma}) \otimes DG) \otimes G(\delta m,\sigma)}\end{footnotesize},
\end{equation}where $\epsilon(m)$ is the energy-dependent detection efficiency determined by MC simulation with a PHSP model, $E_{\gamma}$ is the energy of the M1 transition photon, and $BW(m)$ is the non-relativisic Breit-Wigner ($BW$) function, in which the $\eta_{c}(2S)$ mass and width are fixed to those in the PDG~\cite{Ref_02}. The function $f_{d}(E_{\gamma})$ is used to damp the diverging tail caused by the $E_{\gamma}^{3}$ term, and is given by (${E_{0}^{2}} / {E_{\gamma}E_{0}+(E_{\gamma}-E_{0})^{2}}$), as introduced by the KEDR experiment~\cite{Ref_13}.  Here, $E_{0}$ = $({m^{2}_{\psi(3686)}-m^{2}_{\eta_{c}(2S)}}) / {2m_{\psi(3686)}}$ denotes the peaking energy of the transition photon; $\otimes$ is the convolution operator; $DG$ is a double Gaussian function accounting for the detector resolution, with two parameters obtained by comparing the difference in $M_{2(\pi^{+}\pi^{-})\eta}$ before and after going through the detector; and $G(\delta m,\sigma)$ is a Gaussian function accounting for the difference in detector resolution between data and MC simulation, where $\sigma$ in the $\eta_c(2S)$ mass region is fixed to the value obtained from linear extrapolation from the $\chi_{cJ}$ mass region, while $\delta m$ is allowed to float. 

The shape of the contribution from FSR background is fixed to that obtained from the corrected MC-simulated shape, and the number of events is allowed to float. The contribution from continuum background is fixed to that obtained from off-resonance data and is normalized according to the luminosity and shifted according to the CM energy. The remaining background distribution from $\psi(3686)$ decay is found to be smooth and described by an ARGUS function~\cite{Ref_arg}. 

Figure~\ref{2sfit} shows the fit result with a goodness of fit $\chi^{2}/ndf = 537/360$, where $ndf$ is the number of degrees of freedom. The insert shows the $\eta_{c}(2S)$ mass region on a linear scale. The numerical results are summarized in Table~\ref{list79}. The statistical significance of the $\eta_{c}(2S)$ signal is determined to be 8.4$\sigma$ by comparing the difference in likelihoods with and without including the $\eta_{c}(2S)$ signal component, taking into account the difference in $ndf$ ($\Delta ndf$=2). 

The BFs for our studied channels are calculated via:
\begin{footnotesize}
\begin{equation}\label{bf}
\textstyle{\displaystyle\mathcal{B}(X\to 2(\pp)\eta)  = {\frac{N^{\rm obs}}{N^{\rm tot}_{\rm \psi(3686) \times \displaystyle\mathcal{B}_1 \times \displaystyle\mathcal{B}_2 \times \epsilon }}}},
\end{equation}
\end{footnotesize}where $N^{\rm obs}$ is the number of signal events, $N^{\rm tot}_{\psi(3686)}$ is the total number of $\psi(3686)$ events, $\displaystyle\mathcal{B}_1$ and $\displaystyle\mathcal{B}_2$ are the BFs of $\psip \to \gamma X$ and $\eta \to \gamma\gamma$~\cite{Ref_02}, respectively, and $\epsilon$ is the detection efficiency after correcting the helix parameters of charged tracks~\cite{Ref_18}. Table~\ref{list79} lists the numerical results, and table~\ref{listpbf} lists the product BFs of $\mathcal{B}(\psip \to \gamma X) \times \mathcal{B}(X \to 2(\pp)\eta)$.

\begin{table}[htp]
\begin{center}
\caption{The signal yields, detection efficiencies, and the obtained BFs for $\chi_{cJ}/\eta_{c}(2S) \to 2(\pi^{+}\pi^{-})\eta$ decays. The first and second uncertainties are statistical and systematic, respectively.}
\label{list79}
\begin{tabular}{c c c c}
\hline
        Decay  & $N^{\rm obs}$  & $\epsilon~(\%)$ & $\displaystyle\mathcal{B}$ ($\times 10^{-3}$) \\ \hline
        $\eta_{c}(2S) \to 2(\pi^{+}\pi^{-})\eta$ & 565 $\pm$ $30$ & 6.27 & 12.1 $\pm $0.6$\pm $6.2   \\\hline
        $\chi_{c0} \to 2(\pi^{+}\pi^{-})\eta$ & $42530 \pm 240$ & 8.90 & 4.54$ \pm $0.03$ \pm $0.38    \\\hline
        $\chi_{c1} \to 2(\pi^{+}\pi^{-})\eta$ &$78440 \pm 290$ & 9.31 & 8.07$  \pm $0.03$ \pm $0.76   \\\hline
        $\chi_{c2} \to 2(\pi^{+}\pi^{-})\eta$ & $50980 \pm 240$ & 8.62 & 5.81$ \pm $0.03$ \pm $0.56   \\
        
        \hline
\end{tabular}
\end{center}
\end{table}

The $\eta_{c}$ signal yield is obtained by performing an unbinned maximum likelihood fit to the $M_{2(\pi^{+}\pi^{-})\eta}^{\rm 5C}$ distribution~\cite{Ref_05}. The fit function includes the $\eta_{c}$ signal function and four background components: the background from $\psi(3686)\to \pi^{0}2(\pi^{+}\pi^{-})\eta$, continuum background, non-resonant background ($\psi(3686)\to \gamma2(\pi^{+}\pi^{-})\eta$), and other potential backgrounds from $\psi(3686)$ decays estimated using the $\psi(3686)$ inclusive MC sample. Interference between the $\eta_c$ and the non-resonant process is considered. The fit function is expressed as
\begin{equation}\label{1slineshape}
{\textstyle{{(\epsilon(m) \lvert e^{i\phi}E_{\gamma}^{2} BW(m) + \alpha \mathcal{N} \rvert ^{2} E_{\gamma}^{3})} \otimes G + f_{\rm BKG}}},
\end{equation}where the first term is the $\eta_{c}$ signal shape, which is described by the sum of a modified $BW$ function and a coherent amplitude squared, convolved with the resolution function. Here, $\epsilon(m)$ is the mass-dependent efficiency; $\phi$ is the interference phase angle; $E_{\gamma}$ = $({m_{\psi(3686)}^{2}-m^{2}}) /2m_{\psi(3686)}$ is the energy of the radiative photon; $BW(m)$ is a nonrelativistic $BW$ function defined as $\textstyle{\frac{\Gamma/2}{{(m-m_{0})}+i\Gamma/2}}$, where $m_{0}$ and $\Gamma$ are the mass and width of the $\eta_{c}$; and $\alpha$ is the strength of the non-resonant component. The non-resonant PDF, denoted as $\mathcal{N}$, is described by a second order Chebychev function. The Gaussian function $G$ accounts for the detector resolution, and $f_{\rm BKG}$ denotes the sum of the other three background components mentioned earlier, excluding the non-resonant background. In the fit, $\phi$ and $\alpha$ are free parameters.

The mass-dependent efficiency is determined from MC simulations of the PHSP decay $\eta_{c} \to 2(\pi^{+}\pi^{-})\eta$. We use a Gaussian function to describe the discrepancy between data and MC simulation, and the parameters are determined by fitting the $\jpsi$ peak in the control sample $\psi(3686) \to \gamma\gamma\jpsi, \jpsi \to 2(\pi^{+}\pi^{-})\eta$.

\begin{figure}[htb]
\begin{center}
\includegraphics[width=.4\textwidth]{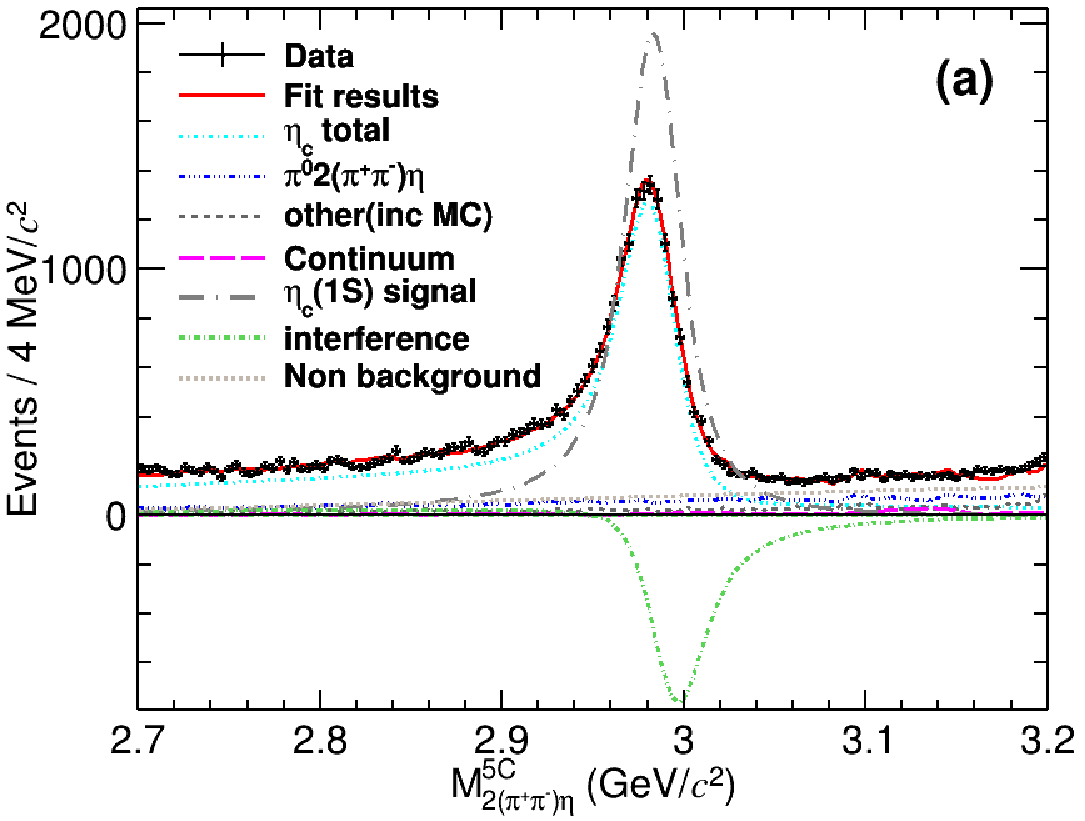}
\includegraphics[width=.4\textwidth]{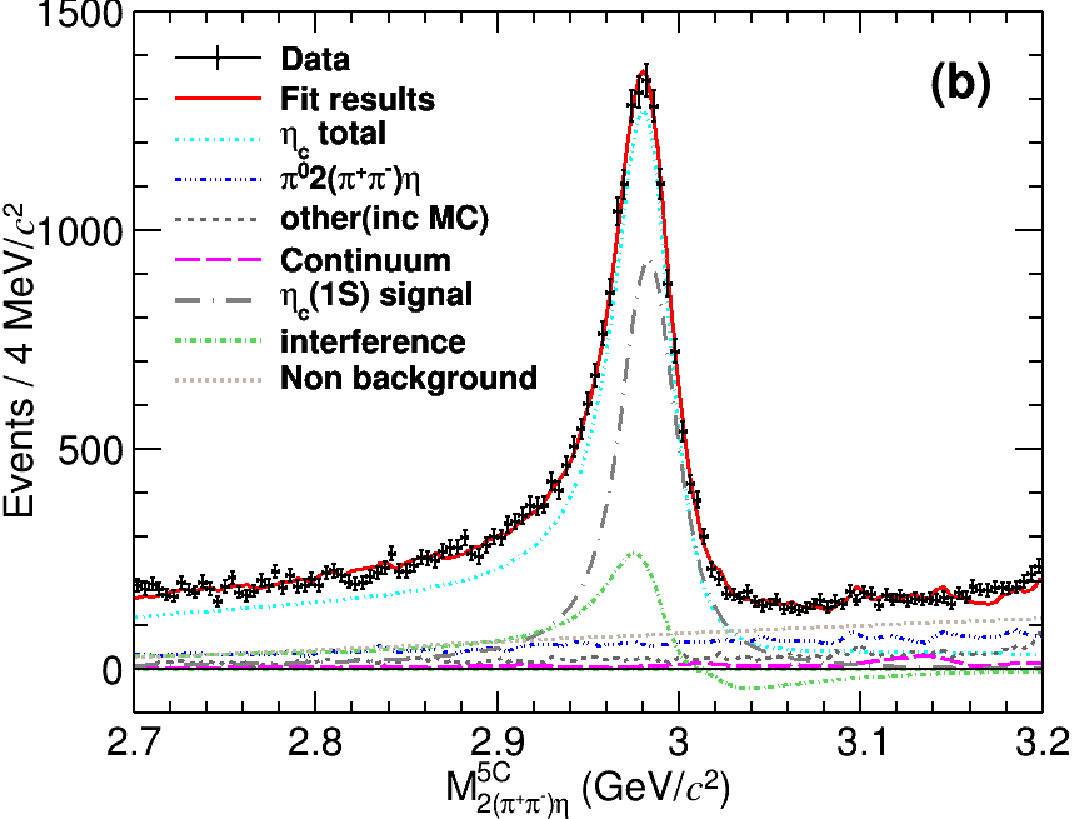}
\caption{The fit results for the $\eta_c$ signal with destructive (a) and constructive (b) interference solutions. The different components are expressed with different line styles or colors.}
\label{1sfit}
\end{center}
\end{figure}

Figure~\ref{1sfit} shows the fit result, where the goodness of fit is $\chi^{2}/ndf = 164/118$. The total number of $\eta_c$ signal is determined to be $N^{\rm obs} = (2.60 \pm 0.24)\times 10^4$ (destructive), or $N^{\rm obs} = (1.24 \pm 0.20)\times 10^4$ (constructive). Using the $\eta_c$ fitted signal yield, we determine the BF of $\eta_c \to 2(\pp)\eta$, listed in Table~\ref{list1s12}, as well as the mass and width of the $\eta_c$ under the two different interference scenarios.

\begin{table}[htb]
\begin{center}
\caption{The obtained mass, width, and BF of the $\eta_{c}$ for the two interference solutions. The first and second uncertainties are statistical and systematic, respectively.}
\label{list1s12}
\scalebox{0.9}{
\begin{tabular}{c c c}
\hline

       Solution & Destructive & Constructive  \\ \hline
        Mass (MeV/$c^{2}$) & $2984.14 \pm 0.12 \pm 0.34$ & $2984.13 \pm 0.13\pm 0.38$   \\ \hline
        Width (MeV) & $28.82 \pm 0.11\pm 0.81$ & $28.81 \pm 0.11\pm 0.82$ \\\hline
        $\mathcal{B}$~($\%$)  & $5.9 \pm 0.5 \pm 2.1$ &  $2.8 \pm 0.5 \pm 1.3$ \\\hline
        $\alpha(\%)$ & $2.541 \pm 0.023$ & $3.50 \pm 0.18$ \\\hline
        $\phi$(rad)  & $-1.949 \pm 0.014$ &  $2.270 \pm 0.030$ \\

        \hline
\end{tabular}
}
\end{center}
\end{table}

\begin{table}[htp]
\begin{center}
\caption{The product BFs of $\mathcal{B}(\psip \to \gamma X) \times \mathcal{B}(X \to 2(\pp)\eta)$. The first and second uncertainties are statistical and systematic, respectively.}
\label{listpbf}
\scalebox{0.9}{
\begin{tabular}{c c}
\hline
        Decay & $\displaystyle\mathcal{B} (\times 10^{-5})$\\ \hline
        $\psip \to \gamma \eta_{c}(2S); \eta_{c}(2S) \to 2(\pi^{+}\pi^{-})\eta$ & $0.85 \pm 0.04 \pm 0.13$    \\\hline
        $\psip \to \gamma \chi_{c0}; \chi_{c0} \to 2(\pi^{+}\pi^{-})\eta$  & $44.4 \pm 0.3 \pm 3.7$    \\\hline
        $\psip \to \gamma \chi_{c1}; \chi_{c1} \to 2(\pi^{+}\pi^{-})\eta$  & $78.7 \pm 0.3 \pm 7.4$   \\\hline
        $\psip \to \gamma \chi_{c2}; \chi_{c2} \to 2(\pi^{+}\pi^{-})\eta$  & $55.3 \pm 0.3 \pm 5.3$   \\ \hline
        $\psip \to \gamma \eta_{c}; \eta_{c} \to 2(\pi^{+}\pi^{-})\eta$ (destructive) & $2.01 \pm 0.17 \pm 0.64$  \\ \hline
        $\psip \to \gamma \eta_{c}; \eta_{c} \to 2(\pi^{+}\pi^{-})\eta$ (constructive) & $0.95 \pm 0.17 \pm 0.44$  \\        
        
        \hline
\end{tabular}
}
\end{center}
\end{table}

The systematic uncertainties related to the measurements for ${\cal B}(\eta_{c}(2S)/\chi_{cJ} \to 2(\pi^{+}\pi^{-})\eta)$ include the following sources. The tracking and PID efficiencies are all assigned to be 1.0$\%$ per charged track~\cite{Ref_14}, and the photon efficiency is also assigned to be 1.0$\%$ per photon~\cite{Ref_15}. The uncertainties caused by the kinematic fit are determined to be (5.5-6.9)$\%$ by comparing the difference in detection efficiency with and without correcting the helix parameters of charged tracks in MC simulation~\cite{Ref_18}. The uncertainties caused by the background veto are determined to be (0.3-6.1)$\%$ by randomly sampling multiple mass windows~\cite{Ref_19}. The uncertainties caused by the number of continuum background events are determined to be 0.23$\%$ for the $\eta_{c}(2S)$ signal and negligible for $\chi_{cJ}$ signals. They are estimated by varying the number of events in the fit by one standard deviation assuming they follow either a Poisson or binomial distribution. The uncertainties caused by MC imperfections are estimated to be (0.1-1.3)\% by comparing the differences with and without correcting the momentum distributions in MC simulation based on data. The uncertainty related to $\mathcal{B}(\eta \to \gamma\gamma)$ is taken from the PDG~\cite{Ref_02}, and the uncertainty related to $\mathcal{B}(\psi(3686) \to \gamma\eta_{c}(2S))$ is taken from Ref.~\cite{Ref_11}. The uncertainty due to the total number of $\psi(3686)$ events is determined to be 0.5$\%$~\cite{Ref_16}. The uncertainties due to MC statistics are determined to be 0.3\% by the same method as for the continuum background events number. The uncertainties caused by the signal yields include the following components: continuum background shape, FSR factor, mass and width of the $\eta_{c}(2S)$, efficiency curve, damping function, detector resolution, and other potential background from $\psi(3686)$ decays. Among them, the uncertainty due to the efficiency curve is estimated by comparing the difference in the fit results with and without including the curve. The uncertainties caused by the continuum and the potential background shapes are estimated by alternatively employing different functions. The uncertainties from other components in signal yields are estimated  by varying the corresponding parameters by one standard deviation. The resulting uncertainties due to signal yields are determined to be (0.2-11.0)$\%$. We assume all these sources are independent and take their sum in quadrature as the total systematic uncertainties, which are determined to be 15.7$\%$ (not including the uncertainty from ${\cal B} (\psi(3686) \to \gamma\eta_{c}(2S))$ measurement), 9.1$\%$, 9.7$\%$, and 9.8$\%$ for $\eta_{c}(2S)/\chi_{cJ}$, respectively.

The uncertainties related to the measurement of ${\cal B}(\eta_{c} \to 2(\pi^{+}\pi^{-})\eta)$ 
have the same or similar sources as above, and are estimated with the same or similar methods. However, the same source may introduce two different uncertainties, corresponding to the constructive and destructive solutions, due to interference between the $\eta_c(1S)$ decay and the non-resonant process. Additionally, the dominant uncertainties come from the non-resonant process with different interference fraction assumptions, which are estimated to be 30.5\% (constructive) or 44.1\% (destructive) by mixing different fractions within one standard deviation, and the largest differences relative to the nominal results are taken as the uncertainties. Finally, the total systematic uncertainty is determined to be 35.2\% (constructive) or 49.1\% (destructive).

The uncertainties related to the measurements of the mass and width of the $\eta_{c}$ are the same as those mentioned above for the $\eta_{c}$ branching fraction measurement and are estimated with the same methods, except for the mass scale, which is estimated using fits to the $\chicJ$ peaks in the same final state. The relative uncertainties are listed in Table \ref{list1s12}.

In summary, the radiative decay $ \psi(3686) \to \gamma2(\pi^{+}\pi^{-})\eta$ is investigated using a sample of 2.7 billion $\psi(3686)$ decays. The BF of $\eta_{c} \to 2(\pi^{+}\pi^{-})\eta$ is determined for the first time under different interference scenarios. Notably, there is a significant difference between the constructive and destructive solutions. The $\chicJ/\eta_{c}(2S) \to 2(\pi^{+}\pi^{-})\eta$ decays are observed for the first time with statistical significance greater than 8$\sigma$, and the corresponding BFs are measured. We present the most precise single measurement of the $\eta_{c}$ mass and width by assuming all radiative non-resonant events interfere with the $\eta_{c}$. The relative phase for constructive or destructive interference is consistent with the earlier BESIII result~\cite{Ref_05} within $1\sigma$, which may suggest a common phase in all modes. The physics behind this possible common phase needs to be understood. With the measured $\eta_c$ mass, we obtain the hyperfine mass splitting to be $\Delta M_{hf}(1S)_{c\bar{c}}= M(\jpsi)-M(\eta_c)=112.8\pm 0.4$ MeV/$c^2$, which is not in conflict with theoretical calculations~\cite{Ref_hf1s1,Ref_hf1s2}.

Furthermore, using the BFs measured in this analysis, the ratio of BFs is
\begin{equation}\label{ratio}
\scriptstyle{ {\frac{\scriptstyle\mathcal{B}(\eta_{c}(2S) \to 2(\pi^{+}\pi^{-})\eta)}{\scriptstyle\mathcal{B}(\eta_{c} \to 2(\pi^{+}\pi^{-})\eta)}} = 
\left\{
\begin{aligned}
0.21 \pm 0.13~(destructive)  \\
0.43 \pm 0.31~(constructive)
\end{aligned}
\right.}.
\end{equation}The results are consistent with most experimental measurements~\cite{Ref_11}, and obviously disagree with the prediction in Ref.~\cite{Ref_10}. The inconsistency between the experimental measurements and the theoretical prediction on the ratio ${\cal B}(\eta_c(2S)\to hadrons)/{\cal B}(\eta_c\to hadrons)$ implies that mixing of the pseudoscalar glueballs to $\eta_{c}$ or $\eta_{c}(2S)$ may play an important role in charmonium decays~\cite{cheny,Ref_mixing1,Ref_mixing2}. Additionally, the distinct contributions of the open-charm loop to the decays of the $\eta_{c}$ and $\eta_{c}(2S)$ cannot be ignored~\cite{loop}. 

The BESIII Collaboration thanks the staff of BEPCII and the IHEP computing center for their strong support. This work is supported in part by National Key R\&D Program of China under Contracts Nos. 2020YFA0406300, 2020YFA0406400, 2023YFA1606000; National Natural Science Foundation of China (NSFC) under Contracts Nos. 11635010, 11735014, 11935015, 11935016, 11935018, 12025502, 12035009, 12035013, 12061131003, 12192260, 12192261, 12192262, 12192263, 12192264, 12192265, 12221005, 12225509, 12235017, 12361141819, 12375070; the Chinese Academy of Sciences (CAS) Large-Scale Scientific Facility Program; the CAS Center for Excellence in Particle Physics (CCEPP); Joint Large-Scale Scientific Facility Funds of the NSFC and CAS under Contract No. U2032108, No. U1832207; 100 Talents Program of CAS; The Institute of Nuclear and Particle Physics (INPAC) and Shanghai Key Laboratory for Particle Physics and Cosmology; German Research Foundation DFG under Contracts Nos. 455635585, FOR5327, GRK 2149; Istituto Nazionale di Fisica Nucleare, Italy; Ministry of Development of Turkey under Contract No. DPT2006K-120470; National Research Foundation of Korea under Contract No. NRF-2022R1A2C1092335; National Science and Technology fund of Mongolia; National Science Research and Innovation Fund (NSRF) via the Program Management Unit for Human Resources $\&$ Institutional Development, Research and Innovation of Thailand under Contract No. B16F640076; Polish National Science Centre under Contract No. 2019/35/O/ST2/02907; The Swedish Research Council; U. S. Department of Energy under Contract No. DE-FG02-05ER41374.

\end{document}

%% file: authors.tex
\author{
  \begin{small}
    \begin{center}
      M.~Ablikim$^{1}$, M.~N.~Achasov$^{13,b}$, P.~Adlarson$^{75}$, R.~Aliberti$^{36}$, A.~Amoroso$^{74A,74C}$, M.~R.~An$^{40}$, Q.~An$^{71,58}$, Y.~Bai$^{57}$, O.~Bakina$^{37}$, I.~Balossino$^{30A}$
% ...
\\
\vspace{0.2cm}
(BESIII Collaboration)\\
\vspace{0.2cm} {\it
$^{1}$ Institute of High Energy Physics, Beijing 100049, People's Republic of China\\
% ...
$^{13}$ GSI Helmholtzcentre for Heavy Ion Research GmbH, D-64291 Darmstadt, Germany\\
% ...
$^{30}$ Inner Mongolia University, Hohhot 010021, People's Republic of China\\
% ...
$^{36}$ Joint Institute for Nuclear Research, 141980 Dubna, Moscow region, Russia\\
$^{37}$ Justus-Liebig-Universitaet Giessen, II. Physikalisches Institut, Heinrich-Buff-Ring 16, D-35392 Giessen, Germany\\
% ...
$^{40}$ Liaoning University, Shenyang 110036, People's Republic of China\\
% ...
$^{57}$ Southeast University, Nanjing 211100, People's Republic of China\\
$^{58}$ State Key Laboratory of Particle Detection and Electronics, Beijing 100049, Hefei 230026, People's Republic of China\\
% ...
$^{71}$ University of Science and Technology of China, Hefei 230026, People's Republic of China\\
% ...
$^{74}$ University of Turin and INFN, (A)University of Turin, I-10125, Turin, Italy; (B)University of Eastern Piedmont, I-15121, Alessandria, Italy; (C)INFN, I-10125, Turin, Italy\\
$^{75}$ Uppsala University, Box 516, SE-75120 Uppsala, Sweden\\
% ...
$^{b}$ Also at the Moscow Institute of Physics and Technology, Moscow 141700, Russia\\
% ...
      }\end{center}
    \vspace{0.4cm}
\end{small}
}
\affiliation{}